\documentclass[a4paper,11pt]{article}
\usepackage{jheppub}
\usepackage{lineno}
\usepackage{comment}
\usepackage{mathtools}
\usepackage{amsthm}
\usepackage{amssymb}
\usepackage{subcaption}
\usepackage{graphicx}
\usepackage{float}
\usepackage{xcolor}
\usepackage{enumitem}
\usepackage{multirow}
\usepackage{physics}
\usepackage{amsmath}
\usepackage{tabularx}
\usepackage{array}
\usepackage{booktabs}
\usepackage{bbold}
\usepackage{hyperref}
\definecolor{customBlue}{RGB}{0,80,160}
\hypersetup{
	colorlinks=true,
	linkcolor=customBlue,
	urlcolor=customBlue,
	citecolor=customBlue
}

\newcolumntype{Y}{>{\raggedright\arraybackslash}X}

\newcommand*\zb{\bar{z}}

\definecolor{colorLY}{rgb}{0,0.6,0}

\makeatletter
\def\@fpheader{\relax}
\makeatother

\title{Conformal Bootstrap Unblocked}

\author{Li-Yuan Chiang, David Poland}
\affiliation{Department of Physics, Yale University, 217 Prospect St, New Haven, CT 06520, USA}

\emailAdd{li-yuan.chiang@yale.edu}
\emailAdd{david.poland@yale.edu}

\abstract{
	We develop a novel conformal bootstrap formulated directly in terms of conformal symmetry, reflection positivity, and crossing symmetry, without explicitly relying on the operator product expansion or conformal blocks. By constructing Gram matrices from states created by pairs of local operators and imposing spectral constraints through conformal generators, we obtain a hierarchy of semidefinite programs that directly constrain correlation functions. We demonstrate the method in one to three dimensions, obtaining correlator and spectral bounds that capture generalized free field behavior and reproduce the characteristic Ising kinks in two and three dimensions. We further show that joint positivity of states at different insertion configurations naturally constrains correlators away from self-dual kinematics.
}

\begin{document}

\maketitle

\pagenumbering{roman}
\setcounter{page}{2}
\pagenumbering{arabic}
\setcounter{page}{1}

\section{Introduction}

The conformal bootstrap has emerged as a powerful non-perturbative approach to constraining conformal field theories (CFTs) through consistency conditions alone. In particular, it has become one of the few methods capable of determining physical observables in strongly coupled theories with rigor and remarkable precision~\cite{Poland:2018epd}. Conformal symmetry organizes local operators into conformal multiplets, whose contributions to correlation functions are encoded in conformal blocks. Together with the convergence of the operator product expansion (OPE)~\cite{Mack:1976pa,Dolan:2000ut,Dolan:2003hv,Pappadopulo:2012jk}, this allows crossing symmetry to be expressed as the equality of conformal block expansions in different channels, reflecting OPE associativity. Unitarity further constrains the spectrum through lower bounds on operator dimensions and the positivity of squared OPE coefficients, both consequences of reflection positivity. These ingredients reduce the conventional bootstrap to constraining the dimensions and OPE coefficients of primary operators.

Since the renaissance of the conformal bootstrap in higher dimensions~\cite{Rattazzi:2008pe}, substantial progress in the computation of conformal blocks~\cite{Kos:2013tga,Erramilli:2020rlr} and numerical optimization~\cite{Simmons-Duffin:2015qma,Landry:2019qug} via semidefinite programming~\cite{Poland:2011ey} has made highly precise bounds on CFT data possible. Particularly intriguing are the sharp kinks and bootstrap islands, which have played an important role in identifying interacting CFTs, most famously the three-dimensional critical Ising model~\cite{El-Showk:2012cjh,El-Showk:2014dwa,Kos:2014bka,Kos:2016ysd,Chang:2024whx}.

While conformal blocks play a central role in the conventional bootstrap, the underlying constraints ultimately originate from more fundamental principles. It is therefore natural to ask whether they are essential or merely one way of implementing these consistency conditions. This raises the central question motivating our work: \emph{can we bootstrap conformal field theories directly from conformal symmetry, crossing symmetry, and reflection positivity, without using conformal blocks?} More specifically, to what extent can these fundamental principles reproduce the constraining power of existing methods, and what new possibilities arise when conformal blocks are no longer the starting point?

Our construction is based on an elementary observation: any finite collection of states created by local operator insertions defines a positive semidefinite Gram matrix, which is a consequence of reflection positivity. Conformal symmetry allows us to express its elements completely in terms of the four-point function and its derivatives with respect to conformally invariant cross ratios, which are further constrained by crossing. In fact, neither the explicit OPE nor the conformal block decomposition is required as an input: crossing symmetry is imposed directly as a relation satisfied by the correlator, while unitarity is implemented through Gram matrix positivity. Spectral assumptions can be incorporated through additional insertions of positive operators constructed from conformal generators. These ingredients lead naturally to a hierarchy of semidefinite programs (SDPs), whose constraints can be systematically strengthened by enlarging the space of probe states.

Our approach is closely related in spirit to the quantum-mechanical bootstrap~\cite{Han:2020bkb,Berenstein:2021dyf,Berenstein:2021loy,Berenstein:2022unr}, where algebraic relations among operators are combined with Gram matrix positivity to constrain physical observables and spectra. More recently, this philosophy has been extended to Euclidean correlation functions~\cite{Cho:2025vws}, where reflection positivity and equations of motion yield semidefinite constraints directly on correlators. The use of positive semidefinite matrices has a much longer history, appearing in the Kac determinant~\cite{Kac:1978ge} and unitarity constraints on conformal representations~\cite{Friedan:1983xq,Mack:1975je}, as well as in reduced-density-matrix methods in quantum chemistry~\cite{PhysRevLett.93.213001,Mazziotti:2012lfg}. The interplay between reflection positivity and crossing symmetry has also been explored in~\cite{Lanosa:2021ods}, where spectral constraints were studied using conformal block expansions. Our approach instead implements reflection positivity and crossing symmetry directly at the level of correlators, without invoking conformal blocks. Moments of CFT data have also been studied extensively in recent bootstrap approaches~\cite{Poland:2024hvb,Chiang:2026nmd,Poland:2026uiy}. In our framework, such moment structures arise naturally as a subset of Gram matrix elements, with the corresponding Hankel matrices appearing as minors.

We develop the general framework in section~\ref{sec:setup}, starting from Hilbert space states prepared by local operator insertions and exploring their positivity constraints through conformal generators. In section~\ref{sec:1d}, we illustrate the construction in $d=1$, where these elementary consistency conditions, together with spectral assumptions, already yield non-trivial bounds on operator dimensions and correlation functions. We then extend the method to higher dimensions in section~\ref{sec:higher_dim}, obtaining bounds that exhibit characteristic Ising kinks in $d=2$ and $d=3$, demonstrating that these elementary consistency conditions already capture highly non-trivial features of interacting CFTs. In section~\ref{sec:multi_kinematics}, we further illustrate how the framework can be applied to directly constrain correlators away from the crossing-symmetric point. Finally, we discuss the limitations of the present approach and possible future extensions in section~\ref{sec:discussion}.

\section{Reflection positive Gram matrices}
\label{sec:setup}

In radial quantization, an inversion in the radius plays the role of conjugation, which, in a unitary theory, defines a positive inner product over the CFT Hilbert space. Our bootstrap is directly based on this idea: we prepare a state by inserting two local primary operators inside a unit sphere, which, when acting on the vacuum, creates a state
\begin{equation}\label{eq:pair_state}
	\ket{\Phi(x, y)} = |x - y|^{2\Delta_\phi} \phi(x) \phi(y) \ket{0}.
\end{equation}
Here, the normalization in front of the state is designed to cancel the two-point correlation function and makes the inner product simply the four-point correlation function. To see this, we begin with the general expression fixed by conformal symmetry:
\begin{equation}
	\langle \phi(x_1) \phi(x_2) \phi(x_3) \phi(x_4) \rangle = \frac{G(u, v)}{x_{12}^{2\Delta_\phi} x_{34}^{2\Delta_\phi}} , \qquad u = \frac{x_{12}^{2}x_{34}^{2}}{x_{13}^{2}x_{24}^{2}}, \quad v = \frac{x_{14}^{2}x_{23}^{2}}{x_{13}^{2}x_{24}^{2}}.
\end{equation}
This solves the conformal Ward identities for identical scalar primaries, leaving an arbitrary function $G(u,v)$ of the cross ratios. For a Hermitian scalar,
\begin{equation}
	\phi(x)^\dagger = |x|^{-2\Delta_\phi} \phi \left( Ix \right), \quad Ix = \frac{x}{|x|^2},
\end{equation}
and we see that
\begin{equation}
	\bra{\Phi(x_3, x_4)} \ket{\Phi(x_1, x_2)} = \frac{x_{12}^{2\Delta_\phi} x_{34}^{2\Delta_\phi}}{|x_3|^{2\Delta_\phi} |x_4|^{2\Delta_\phi}} \langle \phi(x_1)\phi(x_2) \phi(Ix_3) \phi(Ix_4) \rangle = G(\tilde{u}, \tilde{v}),
\end{equation}
where $\tilde{u}$ and $\tilde{v}$ are the conformally invariant cross ratios associated with the configuration $(x_1, x_2, Ix_3, Ix_4)$. We could also consider the state with the vacuum removed,
\begin{equation}
	\ket{\Psi(x, y)} = |x - y|^{2\Delta_\phi} \phi(x) \phi(y) \ket{0} - \ket{0},
\end{equation}
which then gives us the correlator without the identity:
\begin{equation}
	\bra{\Psi(x_3, x_4)} \ket{\Psi(x_1, x_2)} = G(\tilde{u}, \tilde{v}) - 1.
\end{equation}

Now, suppose we have a collection of ``probe states'' that spans a subspace of the Hilbert space, which we denote as
\begin{equation}
	\mathcal{V} = \text{span} \{ \ket{v_1}, \ket{v_2}, ... \} \subset \mathcal{H},
\end{equation}
then the Gram matrix formed by these basis vectors is always positive semidefinite:
\begin{equation}\label{eq:Gram}
	G_{ij} = \bra{v_i} \ket{v_j} \succeq 0,
\end{equation}
which is a direct implication of reflection positivity. Indeed, for any vector $\alpha$,
\begin{equation}
	\alpha^\dagger G \alpha = \sum_{i,j} \alpha_i^* G_{ij} \alpha_j = \left\| \sum_i \alpha_i \ket{v_i} \right\| ^2 \geq 0 \quad \Leftrightarrow \quad G \succeq 0.
\end{equation}
Notice that the positivity of the Gram matrix is preserved under arbitrary changes of basis of $\mathcal{V}$, a freedom that we exploit in later discussions. This observation gives us considerable freedom in choosing the probe states, and by enlarging the probe space $\mathcal{V}$, reflection positivity can be tested on increasingly large subspaces of the Hilbert space.

There are several natural ways to enrich the probe space $\mathcal{V}$:
\begin{enumerate}[label=(\alph*)]
	\item act with conformal generators on \eqref{eq:pair_state}, \label{method:generators}
	\item vary the operator insertion points in \eqref{eq:pair_state}, \label{method:multi_kinematics}
	\item consider states created by non-identical local operators, or
	\item consider states created by more than two local operators.
\end{enumerate}
In this work, we focus on states created by two identical local operator insertions, whose inner products are entirely determined by the corresponding four-point function. We begin by exploring the action of conformal generators, as described in \ref{method:generators}, and later extend our analysis to multiple insertion configurations, i.e.~\ref{method:multi_kinematics}, in section~\ref{sec:multi_kinematics}. The remaining two possibilities naturally generalize our framework to mixed-correlator and higher-point bootstrap problems, which we leave for future work.

Here, we consider probe spaces constructed using ``words'' of conformal generators up to length $L$:
\begin{equation}\label{eq:probe_space}
	\mathcal{V}_L = \text{span}\left\{
	P_{\mu_1}\cdots P_{\mu_j}D^k
	M_{\rho_1\sigma_1}\cdots M_{\rho_l\sigma_l}
	|\Psi(x, y)\rangle:
	\,\, j+k+l \leq L
	\right\}.
\end{equation}
The attentive reader may have noticed the absence of $K_\mu$. The reason is that, at the level of two local operator insertions, any configuration of distinct insertion points can be reached by a rotation followed by a dilatation and a translation. Special conformal generators are therefore redundant in constructing these probe spaces and can be omitted without loss of generality. Moreover, any other ordering of generator insertions can be expressed as linear combinations of states with the ordering above using the conformal algebra. To compute the entries of the corresponding Gram matrix, we apply the conformal Ward identities:
\begin{align}
	[P_\mu,\phi(x)]=\partial_\mu\phi(x), \quad
	& [M_{\mu\nu},\phi(x)]
	=(x_\mu\partial_\nu-x_\nu\partial_\mu)\phi(x), \nonumber \\
	[D,\phi(x)]=(x\cdot\partial+\Delta_\phi)\phi(x), \quad
	& [K_\mu,\phi(x)]=\left(2x_\mu x\cdot\partial-x^2\partial_\mu+2\Delta_\phi x_\mu\right)\phi(x),
\end{align}
which allows us to derive the action of generators on \eqref{eq:pair_state}:
\begin{equation}
	\begin{aligned}
		P_\mu\ket{\Phi(x,y)}
		&= (\partial_{x^\mu}+\partial_{y^\mu})
		\ket{\Phi(x,y)}, \\
		D\ket{\Phi(x,y)}
		&= (x\cdot\partial_x+y\cdot\partial_y)
		\ket{\Phi(x,y)}, \\
		M_{\mu\nu}\ket{\Phi(x,y)}
		&= \bigl(
		x_\mu\partial_{x^\nu}-x_\nu\partial_{x^\mu}
		+y_\mu\partial_{y^\nu}-y_\nu\partial_{y^\mu}
		\bigr)\ket{\Phi(x,y)}, \\
		K_\mu\ket{\Phi(x,y)}
		&= \bigl(
		2x_\mu x\cdot\partial_x-x^2\partial_{x^\mu}
		+2y_\mu y\cdot\partial_y-y^2\partial_{y^\mu}
		\bigr)\ket{\Phi(x,y)}.
	\end{aligned}
\end{equation}
By repeatedly applying these relations, we can systematically compute all entries of the Gram matrix, which are then converted into derivatives of the correlator with respect to the cross ratios,
\begin{equation}
	g_{m,n} = \partial_u^m \partial_v^n G(u, v),
\end{equation}
via a change of variables. Details of this computation can be found in appendix~\ref{appendix:matrix_elements}.

Crossing symmetry follows directly from the permutation invariance of the correlator, together with conformal covariance, which can be implemented without decomposing the correlator using an OPE. In particular, for identical scalar operators, the crossing equation,
\begin{equation}\label{eq:crossing}
	\partial_u^m \partial_v^n \left[ v^{\Delta_\phi} G(u, v) - u^{\Delta_\phi} G(v, u) \right] = 0,
\end{equation}
translates into linear constraints among $g_{m,n}$ that relate entries of the Gram matrix.

Spectral assumptions can be incorporated straightforwardly, although their interpretation requires some care. Suppose that the non-vacuum spectrum in the $\phi \times \phi$ OPE has a gap $\Delta_{\mathrm{gap}}$, then $D-\Delta_{\mathrm{gap}}$ is a positive semidefinite operator over the probe space \eqref{eq:probe_space}, allowing us to construct the gap-localizing Gram matrix
\begin{equation}\label{eq:Gram_gap}
	G^{\text{gap}}_{ij}
	=\langle v_i|(D-\Delta_{\text{gap}})|v_j\rangle \succeq 0.
\end{equation}
Importantly, $\Delta_{\text{gap}}$ constrains the entire non-vacuum spectrum, including both primaries and descendants. In $d=1$, this directly constrains the dimension of the lightest nonidentity primary. In $d\geq2$, the distinction becomes important because spinning primaries also contribute to the spectrum. In the OPE of identical scalars, the unitarity bound requires all spinning primaries to have $\Delta\geq d$, with the stress tensor saturating this bound. Consequently, for $\Delta_{\text{gap}}\leq d$, our constraint is equivalent to imposing a gap on the lightest scalar primary. We will discuss how gap assumptions solely on primary operators can be imposed more generally in section~\ref{sec:higher_dim}.

The familiar unitarity bounds on conformal primaries follow from the positivity of the multiplet Gram matrix, constructed using $P_\mu^\dagger=K_\mu$ and the positivity of $\bra{O_a} K_\mu P_\nu \ket{O^b}$ and $\bra{O_a} K^2 P^2 \ket{O^b}$. In our approach, these constraints are progressively probed through the positivity of the Gram matrices, rather than imposed explicitly as separate bounds on conformal primary dimensions. Rotational covariance also constrains the angular-momentum content of the probe states, and in section~\ref{sec:higher_dim} we further exploit the even-angular-momentum support of antipodal pair states.

The bootstrap constraints described above naturally form a nested hierarchy of semidefinite programs (SDPs) built from the probe spaces
\begin{equation}
	\mathcal{V}_0 \subseteq \mathcal{V}_1 \subseteq \mathcal{V}_2 \subseteq \cdots.
\end{equation}
At each truncation order $L$, we optimize directly over the correlator derivatives, subject to the Gram matrix positivity conditions \eqref{eq:Gram} and \eqref{eq:Gram_gap}, together with the linear crossing equations \eqref{eq:crossing}. Increasing $L$ systematically strengthens these constraints, providing progressively tighter necessary conditions for CFT consistency. In this work, we use \texttt{SDPJSolver.jl}~\cite{sdpj}, an arbitrary-precision semidefinite programming solver developed by one of the authors.

An important question is whether these constraints become sufficient to recover the conventional conformal block formulation in the $L \rightarrow \infty$ limit. We discuss this possibility in section~\ref{sec:discussion}, providing a heuristic argument for how the consistency conditions on operators appearing in the OPE may emerge from Gram matrix positivity.

\section{Examples in one dimension}
\label{sec:1d}

\subsection{A toy example}
\label{subsec:1d_toy}

In this section, we demonstrate our method in $d = 1$, where all the operators are scalar. Let us place the operators at $(x_1, x_2) = (0, x)$, with $0 < x < 1$, creating the state
\begin{equation}
	\ket{\Psi(0, x)} = x^{2\Delta_\phi} \phi(0) \phi(x) \ket{0} - \ket{0}.
\end{equation}
We then consider a special case of \eqref{eq:probe_space}, a subspace where we insert only powers of the dilatation operator, leading to the following matrices:
\begin{equation}
	G^{(0)}=\begin{pmatrix}
		\langle\Psi|\Psi\rangle&\langle\Psi|D|\Psi\rangle\\
		\langle\Psi|D|\Psi\rangle&\langle\Psi|D^2|\Psi\rangle
	\end{pmatrix},\qquad
	G^{(1)}=\begin{pmatrix}
		\langle\Psi|D|\Psi\rangle&\langle\Psi|D^2|\Psi\rangle\\
		\langle\Psi|D^2|\Psi\rangle&\langle\Psi|D^3|\Psi\rangle
	\end{pmatrix},
	\label{eq:radial-example}
\end{equation}
which are exactly the familiar Hankel matrix and shifted Hankel matrix from the Stieltjes moment problem. We then impose
\begin{equation}\label{eq:1d_constraints}
	G^{(0)} \succeq 0, \quad G^{(1)} - \Delta_{\text{gap}} G^{(0)} \succeq 0
\end{equation}
to enforce reflection positivity and a spectral gap. Here, each expectation value is obtained by applying the Ward identity:
\begin{equation}
	D^n \ket{\Psi(0, x)} = (x \partial_x)^n \ket{\Psi(0, x)} \Rightarrow \bra{\Psi(0, y)} D^n \ket{\Psi(0, x)} = (z \partial_z)^n \left( G(z) - 1 \right),
\end{equation}
where $z = x y$ is the cross ratio. To simplify the crossing equation,
\begin{equation}
	(1-z)^{2\Delta_\phi} G(z) = z^{2\Delta_\phi} G(1-z),
\end{equation}
we set $x = 1/\sqrt{2}$, which corresponds to the self-dual kinematics, $z = 1/2$. This fixes all odd derivatives of the correlator, with the leading orders being
\begin{equation}
	g_1 = 4\Delta_\phi g_0, \quad g_3 = 12 \Delta_\phi g_2 + 32 \Delta_\phi (1-4\Delta_\phi^2) g_0, \quad \cdots,
\end{equation}
where $g_{n} = \partial_z^{n}G(z)$ at $z = 1/2$, and the Gram matrices can now be written as
\begin{align}\label{eq:2_by_2_gram}
	G^{(0)} &=
	\begin{pmatrix}
		g_0-1 & 2\Delta_\phi g_0\\
		2\Delta_\phi g_0 & 2\Delta_\phi g_0+\frac14 g_2
	\end{pmatrix}, \nonumber \\
	G^{(1)} &=
	\begin{pmatrix}
		2\Delta_\phi g_0 &
		2\Delta_\phi g_0+\frac14 g_2\\
		2\Delta_\phi g_0+\frac14 g_2 &
		(6\Delta_\phi-16\Delta_\phi^3)g_0
		+\frac34(1+2\Delta_\phi)g_2
	\end{pmatrix},
\end{align}
which already constrain the system in a non-trivial way: \eqref{eq:1d_constraints} and \eqref{eq:2_by_2_gram} imply
\begin{equation}
	\Delta_{\text{gap}} \leq \frac{\bra{\Psi}D\ket{\Psi}}{\bra{\Psi}\ket{\Psi}}  \leq \frac{\bra{\Psi}D^2\ket{\Psi}}{\bra{\Psi}D\ket{\Psi}} \leq 4\Delta_\phi + 3.
\end{equation}
In particular, an upper bound on the gap already emerges from this simple analysis! It would be very interesting to understand whether our analytic result here could be strengthened through a more optimal choice of probe states, closing the gap between the results here and the optimal gap, $2\Delta_\phi + 1$~\cite{Gaiotto:2013nva,Mazac:2016qev,Mazac:2018mdx}.

\subsection{Improving the bounds}

Let us systematically increase the word length, $L$, and include translation generators as well. More precisely:
\begin{equation}
	\mathcal V_L =
	\operatorname{span}\left\{
	P^m D^n |\Psi(0,x)\rangle
	\,\, : \,\,
	m,n\geq 0, \,\, m+n\leq L
	\right\}|_{x = 1/\sqrt{2}}.
\end{equation}
Using the conformal Ward identities, we construct
\begin{equation}
	G^{(0)}_{ij}=\langle v_i|v_j\rangle,
	\qquad
	G^{(1)}_{ij}=\langle v_i|D|v_j\rangle,
\end{equation}
and impose the positivity constraints, \eqref{eq:1d_constraints}. Each matrix has $\sum_{k = 0}^{L} (k+1) = (L+1)(L+2)/2$ rows and columns, and its entries involve correlator derivatives up to order $\Lambda=2L + 1$. After imposing crossing, the independent variables are $\mathbf g=(g_0,g_2,\ldots,g_{2L})$, and both matrices are affine in $\mathbf g$. Therefore, at a fixed assumed gap, we can extremize the correlator by solving the following SDP:
\begin{equation}
	\begin{aligned}
		\text{Extremize} \quad & g_0=G(1/2),\\
		\text{subject to} \quad & G^{(0)} \succeq 0, \quad G^{(1)} - \Delta_{\text{gap}} G^{(0)} \succeq 0.
	\end{aligned}
\end{equation}
To maximize the gap, we perform bisection and look for dual certificates to exclude infeasible gaps, which is explained in detail in appendix~\ref{appendix:feasibility}. The resulting correlator and gap bounds are shown in figures~\ref{fig:1d_bounds} and~\ref{fig:1d_convergence}, respectively. In particular, at the derivative orders studied, our bounds closely track those obtained from the conventional conformal block bootstrap, although they become relatively weak at larger external weights. We will discuss how this could be improved by considering another conformally equivalent kinematic configuration in the section below.

\subsection{Antipodal configuration}
\label{subsec:1d_antipodal}

So far, we have fixed the cross ratio to $z = 1/2$, so that crossing simplifies to linear relations among the correlator derivatives at the same point. However, this corresponds to infinitely many conformally equivalent operator insertion configurations. Although these configurations have the same cross ratio, using the dilatation generator to impose the spectral gap in different frames can yield different constraints at finite truncation.

\begin{figure}[htbp]
	\centering
	\includegraphics[width=\textwidth]{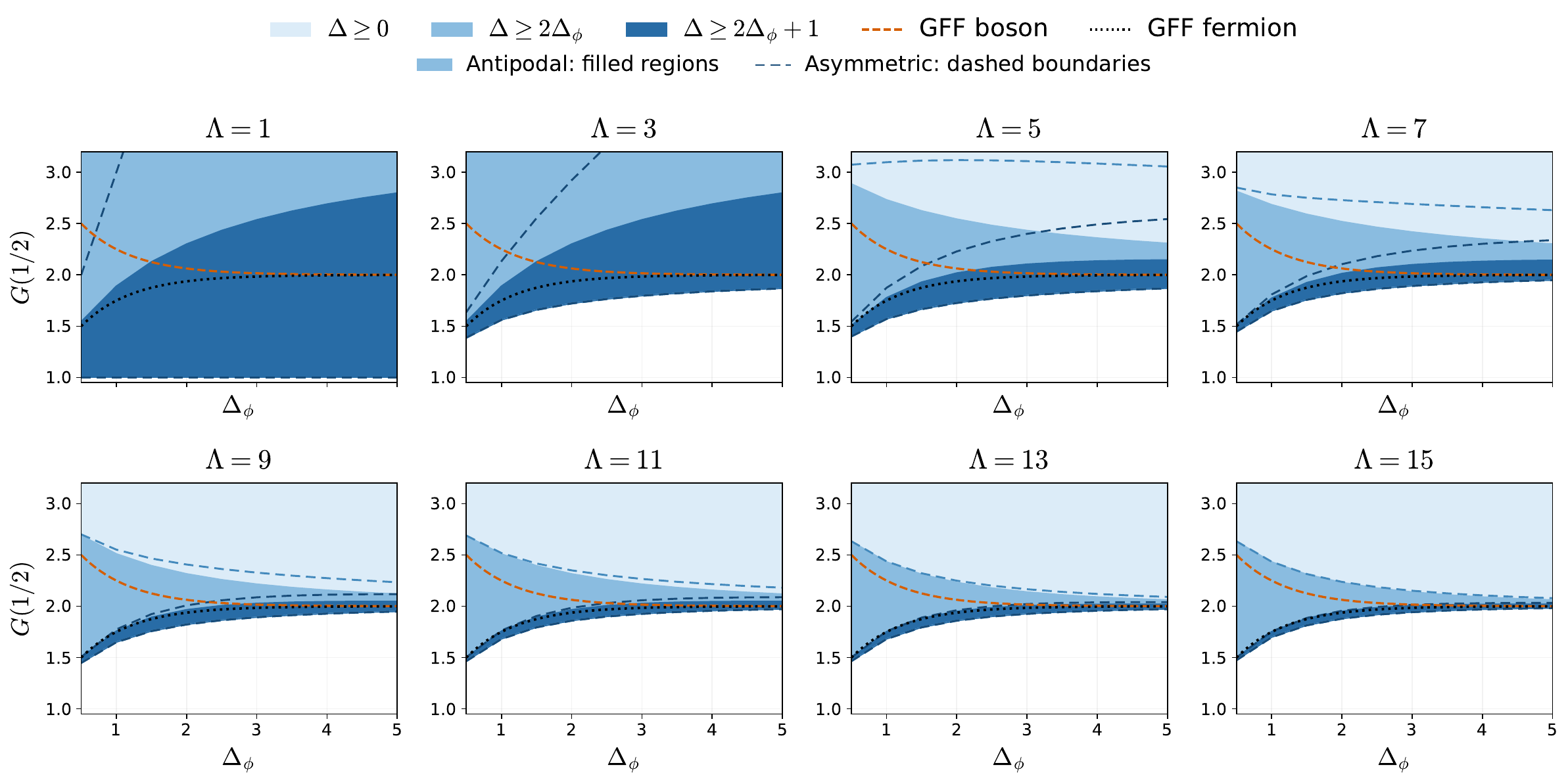}
	\caption{Comparison of correlator bounds at the self-dual point, $G(1/2)$, between the asymmetric (section~\ref{subsec:1d_toy}) and antipodal (section~\ref{subsec:1d_antipodal}) configurations, at increasing derivative orders $\Lambda = 2L+1$. The filled regions represent the bounds obtained in the antipodal frame under different gap assumptions, while the dashed boundaries correspond to those derived from the asymmetric configuration. The orange dashed and black dotted lines represent the generalized free boson and fermion, respectively.}
	\label{fig:1d_bounds}
\end{figure}

\begin{figure}[htbp]
	\centering
	\includegraphics[width=\textwidth]{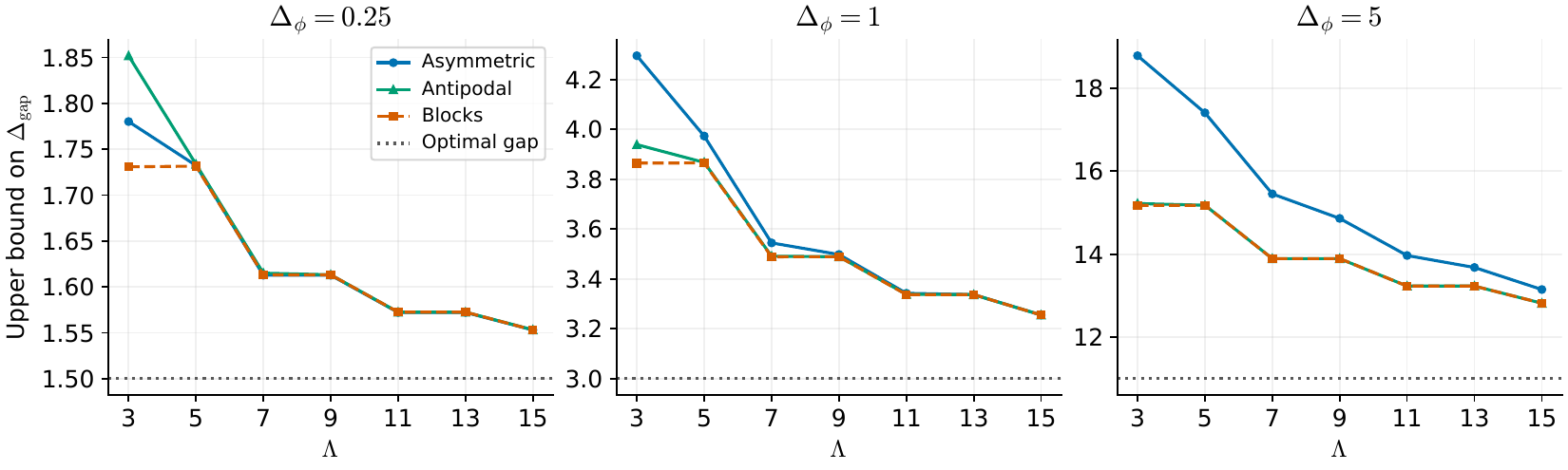}
	\caption{Convergence of the upper bounds on the spectral gap, $\Delta_{\mathrm{gap}}$, as a function of the derivative order $\Lambda$, for $\Delta_\phi = 0.25, 1, \text{and } 5$, in $d=1$. The asymmetric (section~\ref{subsec:1d_toy}) and antipodal (section~\ref{subsec:1d_antipodal}) Gram matrix bounds are compared with those obtained from the conventional conformal block bootstrap and the exact optimal bounds. The antipodal configuration improves convergence particularly at larger external weights, with the bounds nearly coinciding with the conventional bootstrap results at the derivative orders studied.}
	\label{fig:1d_convergence}
\end{figure}

Here, we explore this freedom by considering the same bootstrap problems but with a different insertion configuration, where we place the operators at $(x_1, x_2) = (-s, s)$, where $s = \sqrt{2} - 1$, so that we remain at $z = 1/2$. We call such a kinematic configuration \emph{antipodal}. It turns out that the antipodal configuration, in almost all cases, leads to stronger bounds on both the correlator and the gap, as shown in figure~\ref{fig:1d_bounds} and figure~\ref{fig:1d_convergence}, with most of them matching the bounds on the gap obtained from the conventional bootstrap within the bisection tolerance.

This observation has a natural connection to the radial coordinates introduced in~\cite{Hogervorst:2013sma}. In fact, after a proper rescaling to bring the operators at the conjugate locations to the unit distance from the origin, the antipodal configuration corresponds exactly to the symmetric radial frame, where odd descendant levels vanish and higher-level contributions are strongly suppressed, leading to rapidly convergent conformal block expansions. Intuitively, the antipodal frame would allow us to more easily resolve the primary operators with small Gram matrices, due to the stronger suppression of higher-level descendants. This is in contrast to the $z$-expansion, where the series converges more slowly especially at large exchanged dimensions. Indeed, the configuration considered in section~\ref{subsec:1d_toy} is equivalent, up to an overall rescaling, to the canonical frame with operator insertions at $(0,z,1,\infty)$. Consistent with this observation, the improvement from the antipodal configuration is more significant for large external weights, where we expect heavier operators to dominate the spectrum.

\section{Generalization to higher dimensions}
\label{sec:higher_dim}

Encouraged by these results, we now extend the Gram matrix approach to higher dimensions. We focus on gap maximization in $d=2$ and $d=3$, demonstrating how the characteristic Ising model kinks emerge from our approach.

At first sight, the generalization to $d\geq 2$ seems straightforward. The correlator now depends on two cross ratios, and we can construct Gram matrices by inserting additional conformal generators. One might therefore expect the higher-dimensional bootstrap to follow directly from the same setup. However, there is an important piece of information that we have not yet exploited: \emph{Bose symmetry}.

\subsection{Imposing Bose symmetry}

For identical scalars, the $\phi\times\phi$ OPE contains only even-spin primaries. In the conventional conformal bootstrap, this selection rule is imposed explicitly by restricting the conformal block expansion to even spins. In our approach, however, we do not resolve the correlator into conformal multiplets, and descendants of even-spin primaries can themselves carry odd angular momentum. How, then, can we incorporate this selection rule without introducing conformal blocks?

It turns out that a remarkably simple resolution can be found using the antipodal configuration introduced in section~\ref{subsec:1d_antipodal}. Generalizing to higher dimensions, we consider
\begin{equation}\label{eq:antipodal_pair}
	|\Psi(\mathbf c,\mathbf s)\rangle
	=|2\mathbf s|^{2\Delta_\phi}
	\phi(\mathbf c+\mathbf s)
	\phi(\mathbf c-\mathbf s)|0\rangle-|0\rangle,
\end{equation}
where we eventually set $(\mathbf{c}, \mathbf{s}) =(0, r\hat{n})$, after inserting extra powers of generators and applying Ward identities. Bose symmetry then implies that
\begin{equation}
	|\Psi(0,r\hat n)\rangle
	=|\Psi(0,-r\hat n)\rangle,
\end{equation}
and the odd-spin descendant contributions vanish in this configuration, which is a familiar fact from~\cite{Hogervorst:2013sma}. This can be understood directly from the behavior of the three-point function, $\langle \phi(r\hat n)\phi(-r\hat n)\mathcal O_{\Delta,\ell}(0)\rangle$. Exchanging the two identical scalar insertions sends $\hat n \rightarrow -\hat n$, while rotational covariance implies that performing this picks up a sign of $(-1)^\ell$. Bose symmetry therefore makes the correlator vanish for any odd $\ell$.

This observation allows us to impose additional positivity constraints without explicitly identifying conformal primaries. Let $\mathbf J^2$ denote the quadratic angular Casimir, whose eigenvalues are
\begin{equation}
	\mathbf{J}_\ell^2=\ell(\ell+d-2),
\end{equation}
then the antipodal state has support only at even angular momenta, which implies that the operator
\begin{equation}
	p_\ell(\mathbf{J}^2)=(\mathbf{J}^2-\mathbf{J}^2_\ell)(\mathbf{J}^2-\mathbf{J}^2_{\ell+2}),
	\qquad \ell=0,2,4,\ldots,
\end{equation}
is positive semidefinite on any subspace of $\mathcal{V}_L$ containing only even angular momenta. A sufficient condition is to construct this subspace using operators that commute with the angular Casimir. As a convenient choice, we impose
\begin{equation}\label{eq:bose_constraints}
	\langle w_i|p_\ell(\mathbf{J}^2)|w_j\rangle \succeq0 , \quad
	\langle w_i|p_\ell(\mathbf{J}^2) (D - \Delta_{\text{gap}})|w_j\rangle \succeq0,
\end{equation}
where
\begin{equation}
	|w_i\rangle=D^a \mathbf{J}^{2b}|\Psi(0,r\hat n)\rangle, \quad
	\quad a+2b\leq L-2,
\end{equation}
so that the Gram matrix again involves correlator derivatives up to order $\Lambda = 2L + 1$. This allows us to probe the even-spin selection rule directly through Hilbert space positivity!

The antipodal configuration also provides a convenient coordinate system for constructing the Gram matrices. Writing the two insertion points as $\mathbf c\pm\mathbf s$, the conformal generators act as
\begin{equation}
	\begin{aligned}
		P_\mu|\Psi\rangle
		&=\partial_{c^\mu}|\Psi\rangle,\\
		D|\Psi\rangle
		&=(\mathbf c\cdot\partial_{\mathbf c}
		+\mathbf s\cdot\partial_{\mathbf s})|\Psi\rangle,\\
		M_{\mu\nu}|\Psi\rangle
		&=\bigl(
		c_\mu\partial_{c^\nu}-c_\nu\partial_{c^\mu}
		+s_\mu\partial_{s^\nu}-s_\nu\partial_{s^\mu}
		\bigr)|\Psi\rangle.
	\end{aligned}
\end{equation}
The geometric interpretation is transparent: these generators span all possible infinitesimal variations of the two insertion points. Consequently, the probe space \eqref{eq:probe_space} can equivalently be written as
\begin{equation}\label{eq:probe_space_antipodal}
	\mathcal V_L
	=\operatorname{span}\left\{
	\partial_{\mathbf c}^{\alpha}
	\partial_{\mathbf s}^{\beta}
	|\Psi(\mathbf c,\mathbf s)\rangle
	:
	|\alpha|+|\beta|\leq L
	\right\},
\end{equation}
where $\alpha$ and $\beta$ are $d$-dimensional indices, and we set $\mathbf{c} = 0$ and $|s| = \sqrt{2} - 1$ such that $z = 1/2$, where the correlator derivatives are closed under crossing. This basis also makes counting independent probe states particularly simple: with $2d$ coordinate directions and total derivative order at most $L$, there are
\begin{equation}
	N_L=\binom{2d+L}{L}
\end{equation}
derivative-state labels before symmetry reduction. Since $N_L$ grows rapidly with $L$ and $d$, exploiting the symmetries of the configuration becomes essential. Using selection rules from rotational, parity, and Bose symmetries, states carrying different symmetry quantum numbers have vanishing matrix elements between them, and both the Gram and localizing matrices can be decomposed into smaller independent blocks, without losing any positivity constraints. The method is explained in detail in appendix~\ref{appendix:matrix_elements}.

\begin{figure}[htbp]
	\centering
	\includegraphics[width=\textwidth]{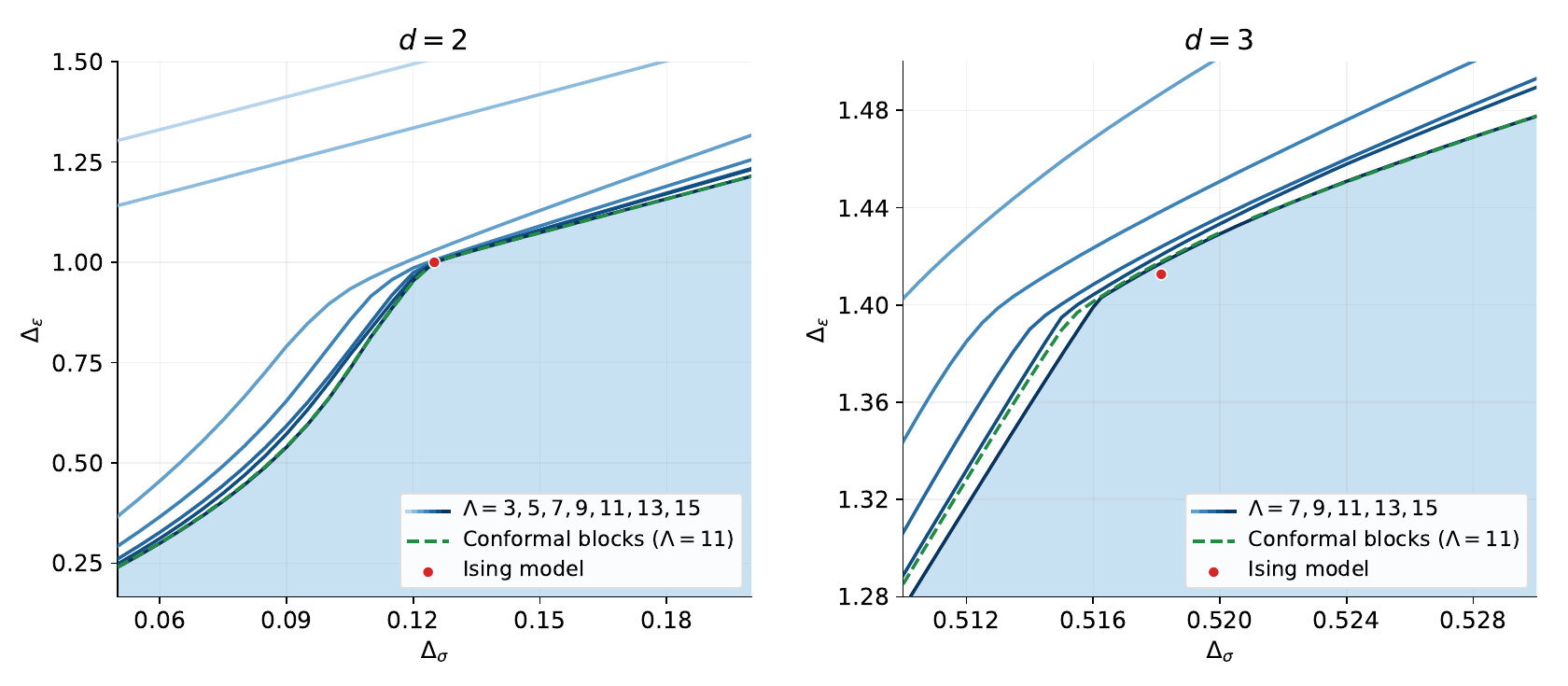}
	\caption{Gap maximization plot around the Ising kinks in $d = 2$ and $d = 3$. After imposing the Bose even-spin constraints, up to spin $\ell = 6$, the bounds significantly improve and converge toward the Ising kinks. The dashed lines represent standard conformal block bootstrap results at similar derivative orders. At a given derivative order, the Gram matrix bootstrap bounds are slightly weaker than those obtained using conformal blocks, but they catch up after a few additional derivatives. Interestingly, the $\Lambda=15$ Gram matrix bounds closely track the $\Lambda=11$ conformal block bounds in $d=2$, and past the kink in $d=3$. The ``gap'' here should be understood as the scaling dimension of the lightest scaling operator in the entire spectrum, including both primaries and descendants, rather than in the usual sense of the lightest scalar primary. However, as argued in section~\ref{sec:setup}, when $\Delta_{\text{gap}} \leq d$, these two notions are essentially equivalent.}
	\label{fig:Ising_kinks}
\end{figure}

We apply the setup above to the critical Ising models in $d=2$ and $d=3$. The numerical results are shown in figure~\ref{fig:Ising_kinks}, where the gap maximization bounds obtained from the Gram matrices converge toward the characteristic Ising kinks as the derivative order increases. Remarkably, imposing the Bose symmetry constraints, \eqref{eq:bose_constraints}, up to only $\ell_{\max}=6$ is already sufficient to resolve the kinks, demonstrating their constraining power. These results show that the Gram matrix approach can reproduce bounds comparable to those obtained from the conventional conformal block bootstrap at similar derivative orders, even though our formulation does not explicitly resolve conformal multiplets.

\subsection{Imposing a second gap}

In~\cite{El-Showk:2012cjh}, the authors considered an additional gap assumption on the subleading $\mathbb{Z}_2$ even scalar primary, $\epsilon'$. Such a gap assumption poses an interesting challenge to our formulation, since the gap-localizing constraint, \eqref{eq:Gram_gap}, constrains all the scaling operators at once, including both primaries and descendants. The question is therefore: how can we impose a gap assumption only on the subleading scalar \emph{primary}, while allowing descendant operators to populate the spectrum below the assumed value?

Here we provide a solution whose essential idea is to construct a non-negative operator whose kernel selects the desired sector, on which we impose the spectral assumptions, with a finite gap in its eigenvalues separating the sector from all other states. To construct such an operator, notice that the primaries are exactly the \emph{kernel} of the non-negative operator
\begin{equation}
	Q = \sum_{\mu} K_\mu^\dagger K_\mu = D(D-d) + \mathbf{J}^2 - \mathcal{C}_2,
\end{equation}
where $\mathcal{C}_2$ is the quadratic conformal Casimir, and the solutions to the equation
\begin{equation}
	\left( Q + \mathbf{J}^2 \right) \ket{\mathcal{O}_{\Delta, \ell = 0}} = 0
\end{equation}
are therefore the scalar primaries. Since both $Q$ and $\mathbf{J}^2$ are positive, we find that only when they are simultaneously zero can the equation be satisfied. Now, we claim that the operator $\mathcal{A} = Q + \mathbf{J}^2$ is exactly the operator with the desired property to isolate the scalar primaries. To see this, we can study the eigenvalues of this operator as follows:
\begin{itemize}
	\item \textbf{Scalar primaries}: $\mathcal{A} = 0$.
	\item \textbf{Descendants in scalar multiplets}: for $\Delta = \Delta_{\mathcal{O}} + n$, the allowed spins are $\ell = n, n-2, \ldots \geq 0$, and we see that
	\begin{align}
		\mathcal{A} &= \Delta(\Delta - d) + 2\ell(\ell + d - 2) - \Delta_{\mathcal{O}}(\Delta_{\mathcal{O}} - d) \nonumber \\
		& \geq \min( 2\Delta_\mathcal{O} + d - 1, 4\Delta_{\mathcal{O}} + 4 - 2d ).
	\end{align}
	\item \textbf{Spinning primaries}: $\mathcal{A} = \mathbf{J}^2 = \ell_{\mathcal{O}}(\ell_{\mathcal{O}} + d - 2) \geq 2d$.
	\item \textbf{Descendants in spinning multiplets}: for $\Delta = \Delta_{\mathcal{O}} + n$, the spin is bounded by $\max(0, \ell_{\mathcal{O}} - n) \leq \ell \leq \ell_{\mathcal{O}} + n$, and we similarly find that
	\begin{align}
		\mathcal{A} &= \Delta(\Delta - d) + 2\ell(\ell + d - 2) - \Delta_{\mathcal{O}}(\Delta_{\mathcal{O}} - d) - \ell_\mathcal{O}(\ell_{\mathcal{O}} + d - 2) \nonumber \\
		& \geq \min(d - 1, 4).
	\end{align}
\end{itemize}
Therefore, for a unitary CFT spectrum with $\Delta_\epsilon>\frac{d-2}{2}$, the operator $\mathcal{A}$ vanishes on scalar primaries, while on all other states it is bounded below by
\begin{equation}
	\mathcal{A} \geq \mathcal{A}_{\min}, \quad \mathcal{A}_{\min} = \min\left( 4\left( \Delta_{\epsilon} - \tfrac{d-2}{2} \right), \,\, d - 1, \,\, 4 \right),
\end{equation}
which is exactly an operator with a finite gap in its spectrum, separating scalar primaries from all other states! We exploit this observation to construct a localizing constraint for the scalar-primary gap:
\begin{equation}\label{eq:filter_first_gap}
	D - \Delta_{\epsilon} + \frac{\max(\Delta_{\epsilon}-E_0, 0)}{\mathcal{A}_{\min}} \mathcal{A} \geq 0,
\end{equation}
where $E_0$ is the gap of the non-scalar-primary spectrum. It can be easily checked that the operator is automatically non-negative on all states that are not scalar primaries, while reducing to $D-\Delta_\epsilon$ on scalar primaries. It therefore imposes $\Delta\geq\Delta_\epsilon$ precisely on scalar primaries, as desired. We can now impose a second gap condition based on the same idea:
\begin{equation}\label{eq:filter_second_gap}
	(D - \Delta_{\epsilon})(D - \Delta_{\epsilon'}) + \frac{(\Delta_{\epsilon'} - \Delta_{\epsilon})^2}{4\mathcal{A}_{\min}} \mathcal{A} \geq 0.
\end{equation}
Again, one could easily verify that this is automatically non-negative on all states that are not scalar primaries, exactly achieving our goal. Gap assumptions on spinning primaries can be incorporated in a similar way.

\begin{figure}[htbp]
	\centering
	\includegraphics[width=\textwidth]{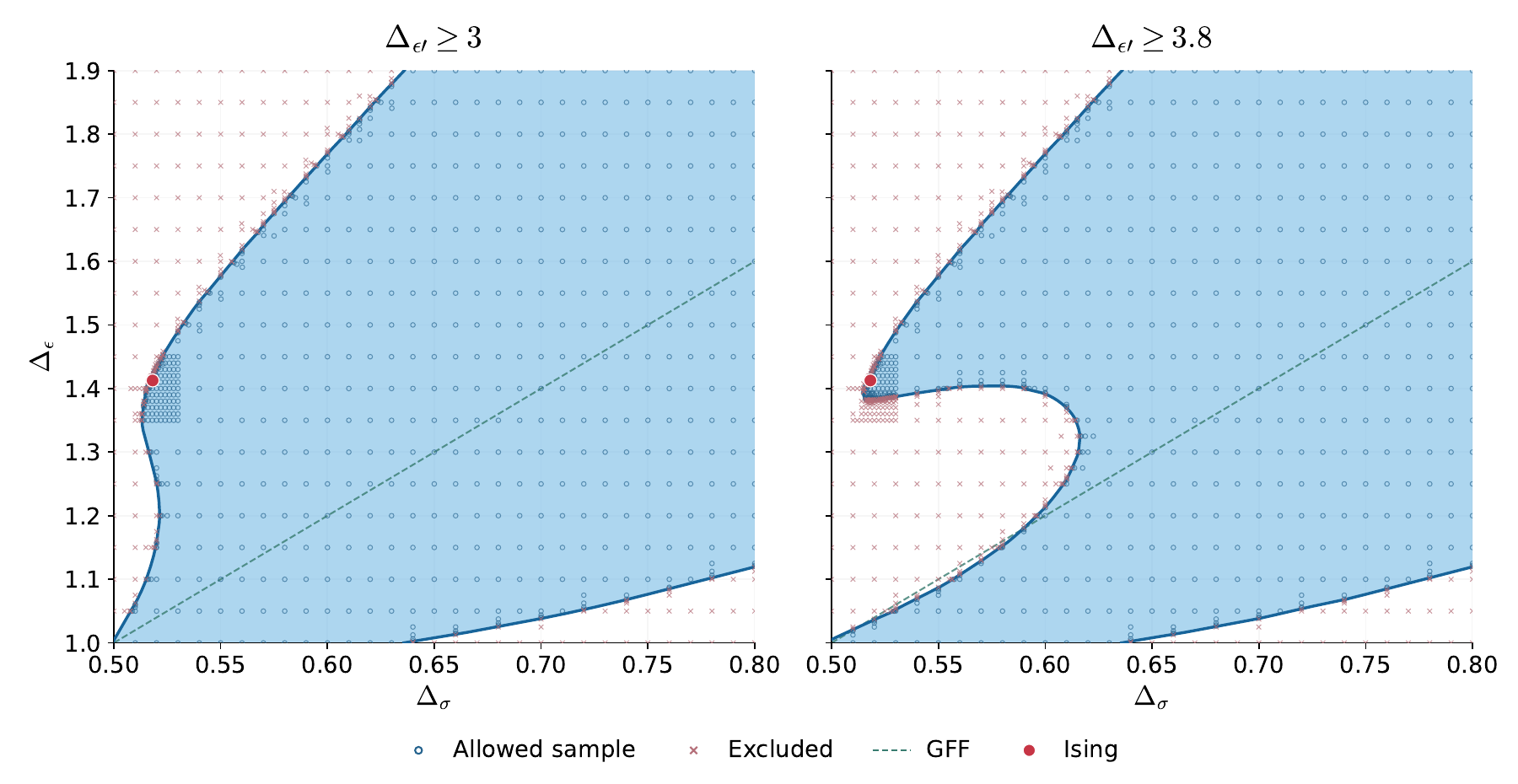}
	\caption{Bounds on the leading scalar dimensions $(\Delta_\sigma,\Delta_\epsilon)$ in $d=3$, imposing additional gap assumptions $\Delta_{\epsilon'}\geq 3$ (left) and $\Delta_{\epsilon'}\geq 3.8$ (right) on the subleading scalar primary, obtained at $\Lambda = 13$. The stronger gap assumption excludes a substantial portion of the allowed region, leaving a narrow peninsula with the Ising model near its tip, reproducing the qualitative behavior observed in~\cite{El-Showk:2012cjh}. The dashed line denotes the generalized free field.}
	\label{fig:Ising_second_gap}
\end{figure}

We apply~\eqref{eq:filter_first_gap} and~\eqref{eq:filter_second_gap} to the 3d Ising model, with results shown in figure~\ref{fig:Ising_second_gap}. A significant portion of the ``continent'' is ruled out, leaving a ``peninsula'' connected to the Ising model and demonstrating that our method can indeed distinguish primaries from descendants.

\subsection{Bounding OPE coefficients}

Although our Gram matrix formulation does not require an OPE expansion to begin with, meaningful information about the OPE coefficients of the low-lying operators can still be extracted. The central idea is that we can \emph{project out} the operator that we wish to study and focus on the spectrum \emph{orthogonal} to that operator. This way, we can systematically write down localizing constraints that hold for the remaining operators.

As an example, we consider the problem of bounding $\lambda_{\sigma \sigma \epsilon}^2$ in the 3d Ising model. The OPE coefficient can be accessed directly through the corresponding three-point function. With the normalization $\langle\epsilon|\epsilon\rangle=1$, we obtain
\begin{equation}
	\bra{\epsilon} \ket{\Psi(x, y)} = \lambda_{\sigma\sigma\epsilon} |x-y|^{\Delta_{\epsilon}}.
\end{equation}
Any inner product $\bra{\epsilon} \ket{v_i}$ can then be computed using Ward identities, which always takes the form
\begin{equation}
	\bra{\epsilon} \ket{v_i} = \lambda_{\sigma \sigma \epsilon} f_i(x, y).
\end{equation}
Let us define probe states with the $\epsilon$ contribution removed,
\begin{equation}
	\ket{w_i} = \ket{v_i} - \bra{\epsilon}\ket{v_i} \ket{\epsilon},
\end{equation}
then the Gram matrix formed by the probe states $\ket{w_i}$ should also be non-negative:
\begin{equation}
	\bra{w_i}\ket{w_j} = G_{ij} - \lambda_{\sigma\sigma\epsilon}^2 f_i^* f_j \succeq 0,
\end{equation}
where we used the fact that $\lambda_{\sigma\sigma\epsilon}\in\mathbb{R}$. This gives $\lambda_{\sigma \sigma \epsilon}^2$ an upper bound. Let us denote the probe space orthogonal to $\ket{\epsilon}$ as
\begin{equation}
	\mathcal{W} = \text{span}\{ \ket{w_1}, \ket{w_2}, ... \}.
\end{equation}
With $\epsilon$ removed, the spectrum now has a gap of $\Delta_{\epsilon'}$ in the scalar primary sector, so we impose that
\begin{equation}
	\mathcal{B} = D - \Delta_{\epsilon'} + (\Delta_{\epsilon'} - E_0) \frac{\mathcal{A}}{\mathcal{A}_{\min}} \succeq 0
\end{equation}
over the space $\mathcal{W}$, where $E_0$ is the gap of the non-scalar-primary spectrum. Here, we estimate conservatively as $E_0 = \Delta_{\epsilon}$ for simplicity. From this, we find that
\begin{equation}
	\bra{w_i} \mathcal{B} \ket{w_j} = \bra{v_i}\mathcal{B}\ket{v_j} + (\Delta_{\epsilon'}-\Delta_\epsilon) \lambda_{\sigma\sigma\epsilon}^2 f_i^*f_j \succeq 0,
\end{equation}
which is crucial for us to derive a lower bound on $\lambda_{\sigma \sigma \epsilon}^2$. Physically, it is intuitive: once we impose a second gap assumption, the $\epsilon$ operator must provide a sufficiently large contribution to the correlator; otherwise, crossing and unitarity cannot be satisfied.

We apply these constraints to the 3d Ising model, fixing $(\Delta_\sigma,\Delta_\epsilon)=(0.518149,1.412625)$ and assuming $\Delta_{\epsilon'}\geq 3.8$. 
We retain the quadratic second-gap constraint~\eqref{eq:filter_second_gap}, testing it and the $\mathcal{B}$-localizing constraint on probes of order at most $L-1$, while the ordinary Gram and dilatation-gap constraints use order $L$. The resulting bounds are summarized in table~\ref{tab:OPE_bounds}. Interestingly, a nontrivial lower bound first emerges at $\Lambda=7$, indicating that sufficiently high-order Gram matrix constraints are needed to resolve the contribution of the $\epsilon$ operator. The bounds then converge rapidly and show good agreement with the known value~\cite{Kos:2016ysd,Simmons-Duffin:2016wlq}, with the implied bounds on $\lambda_{\sigma\sigma\epsilon}$ lying within approximately $0.5\%$ of the reference value at $\Lambda=11$.

More generally, these bounds could be further improved by subtracting additional states, such as descendants of $\epsilon$ or the stress tensor. This would allow the Gram matrices to probe the remaining spectrum more effectively and could also increase the lower bound $\mathcal{A}_{\min}$ on the remaining non-scalar-primary spectrum, strengthening the corresponding spectral constraints. We leave a systematic exploration of this direction to future work.

\begin{table}[htbp]
	\centering
	\begin{tabular}{c|cccc}
		\toprule
		$\Lambda$ & 5 & 7 & 9 & 11 \\
		\midrule
		Lower bound & 0 & 0.95905 & 1.06133 & 1.09451 \\
		Upper bound & 1.42124 & 1.22238 & 1.16946 & 1.11457 \\
		\bottomrule
	\end{tabular}
	\caption{Bounds on $\lambda_{\sigma\sigma\epsilon}^2$ in the 3d Ising model at increasing derivative order $\Lambda$, with $(\Delta_\sigma,\Delta_\epsilon) =(0.518149,1.412625)$ and $\Delta_{\epsilon'}\geq3.8$. The data in \cite{Simmons-Duffin:2016wlq} give $\lambda_{\sigma\sigma\epsilon}^2 = 1.1063962(86)$.}
	\label{tab:OPE_bounds}
\end{table}

\section{Away from the self-dual point}
\label{sec:multi_kinematics}

We now demonstrate another appealing feature of the Gram matrix bootstrap, which is its ability to directly constrain correlation functions away from the self-dual kinematics, where $z = \zb = 1/2$. In the conventional conformal block bootstrap implemented with \texttt{SDPB}~\cite{Simmons-Duffin:2015qma,Landry:2019qug}, extending the analysis to generic kinematic configurations presents several numerical challenges. First, conformal blocks evaluated at different configurations carry different exponential dependence on the exchanged dimensions, which obstructs the polynomial approximation underlying the standard semidefinite programming approach. Second, evaluating conformal blocks becomes increasingly demanding near singular kinematic limits, where the convergence of radial expansions slows down. Finally, contributions from heavy and large-spin operators become increasingly important in these regimes, making numerical truncations less reliable. These difficulties can be partially circumvented using alternative optimization methods. Analytic bounds on correlators away from the crossing-symmetric point were obtained by Paulos~\cite{Paulos:2020zxx}. Paulos and Zheng~\cite{Paulos:2021jxx} obtained numerical bounds using a semi-infinite simplex algorithm~\cite{Paulos:2014vya}. Their implementation however exploits crossing constraints only from derivatives at the self-dual point.

The Gram matrix approach developed in this work offers a different route that, by construction, avoids the problem mentioned above. Moving away from the self-dual point simply requires additional local Taylor data, which are related linearly by crossing symmetry. Moreover, Hilbert space positivity can be imposed not only separately at different configurations, but also jointly on superpositions of states created at different insertion points. As we demonstrate below, these mixed-configuration constraints can significantly strengthen the resulting bounds.

Here, we consider bounding the correlator along the diagonal line $z=\bar z\in(0,1)$ as an example, although the method can be straightforwardly generalized to points away from the diagonal. For any chosen $z$, we introduce two sets of local correlator derivatives at $z$ and $1-z$, related linearly by crossing. We can therefore construct the corresponding Gram matrices using the same procedure as before for the two different kinematic configurations and then optimize $G(z,z)$. To further exploit Hilbert space positivity, we combine the probe spaces associated with the two configurations.

Let us make this procedure more explicit by recalling the antipodal state $|\Psi(0,r\hat z)\rangle$ introduced in section~\ref{sec:higher_dim}, where $r$ is related to the cross ratio via~\cite{Hogervorst:2013sma}
\begin{equation}
	z=\frac{4r^2}{(1+r^2)^2}.
\end{equation}
Now, crossing relates two points with different radial coordinates, $r_1=r(z)$ and $r_2=r(1-z)$, and we construct the joint probe space as
\begin{equation}
	\tilde{\mathcal V}_L
	=\mathcal V_L^{(r_1)} + \mathcal V_L^{(r_2)},
	\label{eq:joint_probe}
\end{equation}
where $\mathcal V_L^{(r)}$ denotes the space \eqref{eq:probe_space_antipodal} evaluated at $\mathbf s=r\hat z$. This allows us to impose reflection positivity on arbitrary superpositions of states created at the two configurations, through the joint Gram matrix
\begin{equation}\label{eq:joint_gram}
	\tilde G=
	\begin{pmatrix}
		G^{(11)} & G^{(12)}\\
		G^{(21)} & G^{(22)}
	\end{pmatrix}\succeq0,
	\qquad
	G^{(ij)}_{ab}
	=\langle v_a^{(r_i)}|v_b^{(r_j)}\rangle.
\end{equation}
We similarly impose the gap-localizing matrix
\begin{equation}\label{eq:joint_gram_gap}
	\tilde G^{\mathrm{gap}}=
	\begin{pmatrix}
		G^{\mathrm{gap},(11)} &
		G^{\mathrm{gap},(12)}\\
		G^{\mathrm{gap},(21)} &
		G^{\mathrm{gap},(22)}
	\end{pmatrix}\succeq0,
	\qquad
	G^{\mathrm{gap},(ij)}_{ab}
	=\langle v_a^{(r_i)}|
	(D-\Delta_{\mathrm{gap}})
	|v_b^{(r_j)}\rangle.
\end{equation}
Again, Bose symmetry is imposed through the quadratic spin Casimir localizing constraint introduced in section~\ref{sec:higher_dim}, now as a joint positive condition as well. Here, the off-diagonal blocks, which consist of inner products formed between elements of probe spaces at different kinematics, involve correlator derivatives at an additional diagonal configuration, with
\begin{equation}
	z'=\frac{4r_1r_2}{(1+r_1r_2)^2}.
\end{equation}
This requires us to introduce another set of local Taylor data at $z'$, related to those at $1-z'$ by crossing symmetry, together with the corresponding positivity constraints:
\begin{equation}
	G^{(33)} \succeq 0, \,\, G^{(44)} \succeq 0, \qquad r_3 = r(z'), \,\, r_4 = r(1 - z').
\end{equation}
At each of these additional configurations, we also impose the gap-localizing constraint and the Bose constraints~\eqref{eq:bose_constraints}, with the same derivative truncation.
We see that imposing joint positivity at two crossing-related configurations naturally generates additional crossing and positivity constraints at new kinematic points! This observation suggests a recursive generalization of our construction: the newly generated configurations can themselves be combined into larger joint probe spaces, whose mixed Gram matrix elements probe yet further insertion geometries, generating an expanding hierarchy of constraints across different kinematic configurations. We leave a systematic exploration of this recursive procedure for future work, and restrict ourselves to the first iteration in the present analysis.

\begin{figure}[htpb]
	\centering
	\includegraphics[width=\textwidth]{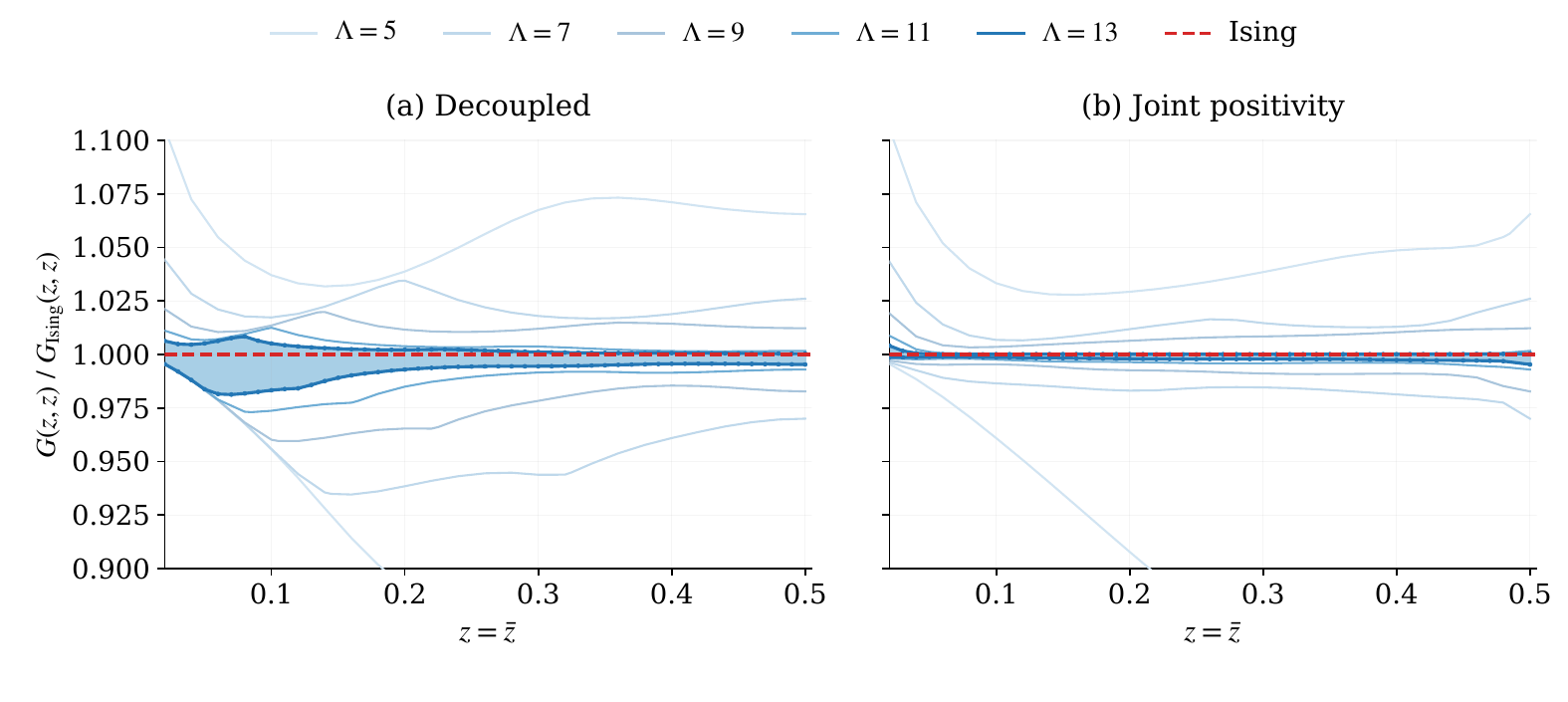}
	\caption{Bounds on $G(z,z)$ in $d=3$ with $\Delta_\phi=0.518149$, $\Delta_{\mathrm{gap}}=1.412625$, normalized by the Ising correlator computed with the OPE data from~\cite{Simmons-Duffin:2016wlq}. We compare bounds obtained by separately imposing reflection positivity and those obtained by joint positivity. The latter significantly tightens the bounds and improves convergence, constraining the correlator to within $1\%$ of the Ising correlator estimate over the displayed range at $\Lambda=13$.}
	\label{fig:joint_vs_decoupled_bounds}
\end{figure}

We test this method on the 3d Ising model, where we bound correlators at various different cross ratios with the following parameters:
\begin{equation}
	d = 3, \quad \Delta_\phi=0.518149, \quad \Delta_{\mathrm{gap}}=1.412625, \quad \Lambda = 13,
\end{equation}
and compare bounds obtained using decoupled positivity, i.e. $G^{(11)}, G^{(22)} \succeq 0$, and joint positivity, i.e. $\tilde{G} \succeq 0$. We find that joint positivity leads to substantially improved bounds, as shown in figure~\ref{fig:joint_vs_decoupled_bounds}. We also compare the bounds with an estimate computed from the conformal data of~\cite{Simmons-Duffin:2016wlq} by summing the contributions of all ``stable operators'' in the extremal spectrum. Away from the Euclidean diagonal, correlator values enter as overlaps between pair states with different orientations. Such overlaps can likewise be constrained by joint Gram positivity, together with crossing symmetry. We leave a numerical implementation of this extension to future work.

\section{Discussion}
\label{sec:discussion}

In this work, we have developed an alternative approach to the conformal bootstrap, formulated directly in terms of conformal symmetry, crossing symmetry, and Hilbert space positivity. Remarkably, without explicitly introducing the OPE or conformal blocks, we obtain non-trivial bounds on operator dimensions and correlation functions, including the characteristic Ising kinks in two and three dimensions. The framework also shows flexibility in constraining correlators away from the crossing-symmetric point, probing crossing and unitarity more globally. These results suggest that much of the constraining power of the conventional conformal bootstrap can be accessed directly through these fundamental consistency conditions.

Despite these encouraging results, the present implementation remains less efficient than conventional methods. In particular, the size of the Gram matrices grows rapidly with the derivative order, making higher-order computations expensive. Improving numerical efficiency is therefore an important direction for future work. Possible strategies include exploiting sparsity and low-rank structure in the SDP more aggressively, exploring the possibilities of lower-precision computations through proper numerical conditioning, as well as identifying more efficient choices of probe states. Indeed, the effectiveness of our approach depends crucially on which directions in the Hilbert space are tested, and the probe spaces considered here are not meant to be claimed as optimal.

An important question is whether the Gram matrix bootstrap becomes equivalent to the conventional conformal block bootstrap in the infinite-truncation limit, as $L\rightarrow\infty$. One direction follows straightforwardly from the conventional formulation: expanding the pair-insertion states in an orthonormal basis of unitary conformal multiplets, the Gram matrix takes the form
$$
	G_{ij}=\sum_{\mathcal O}p_{\mathcal O}\sum_\alpha
	C_{\mathcal O,\alpha}^{(i)*}C_{\mathcal O,\alpha}^{(j)},
	\qquad p_{\mathcal O}\geq0,
$$
where $\alpha$ labels descendants and $C_{\mathcal O,\alpha}^{(i)}$ are fixed by conformal symmetry. Thus, a positive conformal block expansion guarantees $G[\mathcal V_L]\succeq0$ at every truncation order. The converse is, however, more subtle. Since our construction already incorporates conformal covariance and the radial adjoint structure, the question is whether the infinite limit is sufficiently complete to resolve the full spectrum of scaling operators. A possible route toward equivalence is through spectral filtering. Assuming a discrete, lower-bounded dilatation spectrum, radial evolution isolates the lowest energy component of the probe state,
$$
	\ket{E_0} \propto \lim_{t\to-\infty}e^{-E_0t}e^{tD}|\Psi(x, y)\rangle,
$$
and states with higher energies can be extracted successively. If these spectral filters can be approximated within the infinite probe space, Gram matrix positivity would extend to the reconstructed scaling eigenstates. Together with the conformal algebra, which organizes these states into unitary conformal multiplets, this suggests that the usual unitarity bounds and positive expansion coefficients could be recovered. This equivalence question may also be related to the Osterwalder–Schrader reconstruction theorem~\cite{Osterwalder:1973dx}, which connects reflection positivity to the reconstruction of a unitary Hilbert space. Furthermore, certain submatrices of our Gram matrices take the form of moment matrices, suggesting potential connections to the moment problem and recent studies of moments in the conformal bootstrap~\cite{Chiang:2026nmd,Poland:2024hvb,Poland:2026uiy}. A rigorous proof of equivalence and a deeper exploration of these connections are left for future work.

The flexibility of our framework also opens several directions for future development, particularly for bootstrap problems where the explicit computation of conformal blocks becomes challenging. A natural generalization is to consider states created by non-identical or spinning local operators, leading to mixed-correlator and spinning-correlator bootstrap. Excitingly, the same philosophy extends naturally to higher-point correlation functions, where the positivity of states created by multiple operator insertions provides additional consistency conditions and, more importantly, allows us to probe the consistency between lower- and higher-point correlators. Related developments in the six-point bootstrap have shown that positive semidefinite constraints can be formulated in descendant space without explicitly evaluating six-point conformal blocks~\cite{Antunes:2023kyz,Harris:2025cxx}. We are currently extending our formulation to higher-point correlation functions, with preliminary results supporting the applicability of this approach to six-point functions~\cite{Chiang:6ptWIP}. For example, using a basis constructed at $L=6$ acting on $\ket{\sigma}$ and $\ket{\sigma \sigma \sigma}$, and incorporating known 3d Ising four-point data, we are able to bound the subleading $\mathbb{Z}_2$-odd gap as $\Delta_{\textrm{odd}} \lesssim 4.25$ and constrain the normalized hexagon-configuration six-point function $\langle\sigma\sigma\sigma\sigma\sigma\sigma\rangle/\langle\sigma^2\rangle^3$ to a narrow window $\sim \left[5.34, 5.35\right]$. We take this as a promising sign of the power of this approach to extract information from higher-point functions.

On the other hand, our method has already demonstrated that in $d=1$, even a $2\times 2$ Gram matrix can yield nontrivial bounds on the spectral gap. It is therefore interesting to explore connections with the analytical conformal bootstrap, where suitable choices of probe states, perhaps via smearing over operator insertion locations, may provide new ways to derive analytical constraints on CFT data. Another promising direction is to extend our approach to Lorentzian kinematics, with potential applications to the lightcone bootstrap~\cite{Fitzpatrick:2012yx,Komargodski:2012ek,Caron-Huot:2017vep}, Regge physics~\cite{Costa:2012cb}, and chaos bounds~\cite{Maldacena:2015waa}. In particular, positive semidefinite matrices of averaged null energy expectation values have been used to derive constraints on CFT data~\cite{Hofman:2008ar,Hartman:2016lgu,Cordova:2017zej,Cordova:2017dhq}, suggesting a natural connection to our Gram matrix framework.

Finally, it would be desirable to extract information on the CFT data directly from Gram matrices, potentially revealing interesting connections with the extremal functional method~\cite{El-Showk:2012vjm} and the literature on moment problems. For instance, in the Wilson–Fisher fixed points in $d=4-\epsilon$, multiplet recombination follows from the equations of motion~\cite{Rychkov:2015naa}, while previous bootstrap studies have found evidence of operator decoupling near the Ising kink~\cite{El-Showk:2014dwa}. From a complementary perspective, extremal Gram matrices can become nearly singular, with their null vectors corresponding to linear combinations of probe states with vanishing norm. Understanding the structure and physical interpretation of these approximately null states may offer new insights into operator decoupling and the underlying operator relations.

In conclusion, our work demonstrates that conformal symmetry, crossing symmetry, and unitarity can be directly utilized in the bootstrap by probing the CFT Hilbert space through Gram matrix positivity. We hope that the approach developed here will provide a useful starting point for further investigations of conformal field theories.

\acknowledgments
We thank Yu-tin Huang, Tony Liu, Gordon Rogelberg, and Naveen Umasankar for valuable discussions. The authors were supported by DOE grant DE-SC0017660. Computations were performed on the Yale Grace and Bouchet computing clusters, supported by the facilities and staff of the Yale University Faculty of Sciences High Performance Computing Center.

\appendix

\section{Computing Gram matrix elements}
\label{appendix:matrix_elements}

Here, we explain how the Gram matrix elements are computed and how the matrix can be reduced into smaller blocks by exploiting the symmetries of the antipodal configuration. Given the probe space defined in \eqref{eq:probe_space_antipodal}, the Gram matrix elements can be obtained by differentiating the overlap $\langle\Psi(\mathbf c',\mathbf s')|\Psi(\mathbf c,\mathbf s)\rangle=G(u, v)-1$ with respect to $\mathbf{c}$ and $\mathbf{s}$. Since conformal covariance fixes the dependence of this overlap on the cross ratios, we can express the matrix elements as
\begin{equation}
	G_{ij}(x_1,x_2,x_3,x_4)
	=\sum_{m,n}A_{ij}^{m,n}(x_1,x_2,x_3,x_4)
	\partial_u^m\partial_v^nG(u,v) - \delta_{i0} \delta_{j0},
\end{equation}
where $A_{ij}^{m,n}$ are coefficients determined by the insertion coordinates, and the second term accounts for the vacuum contribution. For the complete derivative spaces used here, configurations related by a conformal transformation preserving the unit ball have Gram matrices related by an invertible change of probe basis, and therefore impose equivalent positivity constraints. The gap-localizing constraints can nevertheless differ at finite truncation, because the dilatation generator changes under transformations between frames. This accounts for the frame dependence observed in section~\ref{subsec:1d_antipodal}. In practice, it is more convenient to implement crossing symmetry at the self-dual kinematics using the coordinates $a=z+\bar z-1$ and $b=(z-\bar z)^2$, under which crossing acts simply as $a\mapsto-a$. This change of variables results in an invertible linear map on the correlator derivatives at any fixed truncation order, yielding equivalent Gram matrix constraints while making it easier to identify independent variables in the SDP.

The Gram matrix can be further simplified by exploiting the \emph{little group} of the antipodal configuration, which describes the spatial transformations that preserve the insertion points, including their exchange permitted by Bose symmetry. For identical scalars, this group is $O(d-1)\times\mathbb Z_2$, which becomes $\mathbb Z_2\times\mathbb Z_2$ in $d=2$ and $O(2)\times\mathbb Z_2$ in $d=3$. Since these transformations preserve the inner product and commute with both $D$ and the angular Casimir $\mathbf{J}^2$, the Gram matrices and all corresponding localizing matrices can be decomposed into orthogonal symmetry sectors. We demonstrate this reduction in $d=3$, with antipodal insertions $\pm r\hat z$, where the little group consists of the following symmetries:
\begin{enumerate}[label=(\roman*)]
	\item \textbf{Rotations around the $z$-axis:}
	The rotational subgroup $SO(2)\simeq U(1)$ allows us to label probe states by their helicity. A convenient basis is
	\begin{equation}
		\partial_{c^\pm}
		=\frac{1}{\sqrt{2}}
		(\partial_{c^x}\pm i\partial_{c^y}),
		\qquad
		\partial_{s^\pm}
		=\frac{1}{\sqrt{2}}
		(\partial_{s^x}\pm i\partial_{s^y}),
	\end{equation}
	which carry helicities $\pm1$, while derivatives along $z$ carry zero helicity. In this basis, each probe state has a definite total helicity $h$, and rotational invariance implies that matrix elements between states of different helicities vanish.

	\item \textbf{Parity:}
	Under $P: (x,y,z) \mapsto (-x,-y,-z)$, the antipodal coordinates transform as $(\mathbf c,\mathbf s)\mapsto(-\mathbf c,-\mathbf s)$. Bose symmetry, i.e. $|\Psi(\mathbf c,\mathbf s)\rangle=|\Psi(\mathbf c,-\mathbf s)\rangle$, then implies
	\begin{equation}
		P |\Psi(\mathbf c,\mathbf s)\rangle
		=|\Psi(-\mathbf c,\mathbf s)\rangle.
	\end{equation}
	Consequently, the probe states have definite parity
	\begin{equation}
		\eta=(-1)^{|\alpha_c|},
	\end{equation}
	where $|\alpha_c|$ denotes the total number of $\mathbf{c}$-derivatives. States with opposite parity are orthogonal, leading to further vanishing matrix elements.

	\item \textbf{Transverse reflection:}
	Finally, consider $R:(x,y,z)\mapsto(x,-y,z)$, which swaps $\partial_+\leftrightarrow\partial_-$ and hence sends $h$ to $-h$, relating equivalent Gram matrix blocks. The special case is $h=0$, which is closed under this reflection, allowing a decomposition into reflection-even and reflection-odd states.
\end{enumerate}
Together, these selection rules decompose the Gram matrices into independent blocks labeled by $h$ and $\eta$, with an additional reflection parity organization for $h=0$, reducing the size of the matrices in the SDP without weakening them.

\section{SDP feasibility problem}
\label{appendix:feasibility}

At a fixed external scaling dimension and an assumed spectral gap, our goal is to determine whether there exists a set of correlator derivatives satisfying crossing symmetry and all the Gram matrix positivity constraints. Let us write the positivity condition as
\begin{equation}
	G(\mathbf g)=G_0+\sum_a g_a G_a \succeq 0,
\end{equation}
where $G_0$ and $G_a$ are known matrices, and $g_a$ are the remaining independent correlator derivatives after crossing is imposed. To test feasibility, we introduce an auxiliary variable $t$ and solve
\begin{equation}
	\begin{aligned}
		\text{Maximize}\quad & t,\\
		\text{subject to}\quad & G(\mathbf g)-tI\succeq0.
	\end{aligned}
\end{equation}
This effectively maximizes the smallest eigenvalue of $G(\mathbf g)$, and a feasible point should return $t\geq0$, while a strictly negative optimal value rules out the assumed gap. An alternative formulation considers the problem from the dual side:
\begin{equation}
	\begin{aligned}
		\text{Minimize}\quad & \text{Tr}(YG_0),\\
		\text{subject to}\quad
		& Y\succeq0,\qquad \text{Tr}Y=1,\\
		& \text{Tr}(YG_a)=0
		\quad\text{for all }a.
	\end{aligned}
\end{equation}
Any strictly negative objective value then contradicts primal feasibility:
\begin{equation}
	0
	\underbrace{\leq}_{\text{primal feasibility}}
	\text{Tr}[YG(\mathbf g)]
	=\text{Tr}(YG_0)
	+\sum_a g_a\text{Tr}(YG_a)
	\underbrace{=}_{\text{dual feasibility}}
	\text{Tr}(YG_0) < 0.
\end{equation}
The gap maximization bounds presented in this work were obtained using bisection, with excluded gaps certified by dual feasibility. We use \texttt{SDPJSolver.jl} with 256-bit arithmetic and set the duality gap, primal feasibility, and dual feasibility thresholds to $10^{-24}$. The bisection threshold is set to $2 \times 10^{-3}$ for the $d = 1$ bounds and $10^{-4}$ for the $d \geq 2$ bounds. 

\bibliographystyle{JHEP}
\bibliography{biblio.bib}

\end{document}